\documentclass[fleqn,twoside]{article}
\usepackage[headings]{espcrc2}
\readRCS
$Id: espcrc2.tex,v 1.2 2004/02/24 11:22:11 spepping Exp $
\usepackage{graphicx}
\usepackage[figuresright]{rotating}

\newcommand{\AmS}{{\protect\the\textfont2
  A\kern-.1667em\lower.5ex\hbox{M}\kern-.125emS}}

\title{Impact of QED corrections to Higgs decay into four leptons at the LHC}

\author{
C. M. Carloni Calame\address[pvinfn]{INFN, Sezione di Pavia, via A. Bassi, 
6, Pavia (Italy)}\address[pvuniv]{Dipartimento di Fisica Nucleare e Teorica,
Universit\`a di Pavia, via A. Bassi, 6, Pavia (Italy)},
M. Moretti\address[fe]{Dipartimento di Fisica, Universit\`a di Ferrara, via
Saragat, 1, Ferrara (Italy)},
G. Montagna\addressmark[pvuniv]\addressmark[pvinfn],
O. Nicrosini\addressmark[pvinfn]\addressmark[pvuniv],
F. Piccinini\addressmark[pvinfn]\addressmark[pvuniv],
A.D. Polosa\address[ba]{INFN, Sezione di Roma, piazzale A. Moro, 2, Roma
(Italy)}}

\begin{document}
\begin{abstract}
At the LHC a precise measurement of the Higgs boson mass (if discovered), 
at the level of 0.1-1\%, will be possible through the channel $g g\to H\to 4l$ 
for a wide range of Higgs mass values. To match such an accuracy, the
systematic effects induced by QED corrections need to be investigated. In the
present study the calculation of ${\cal O}(\alpha)$ and higher order QED
corrections is illustrated as well as their impact on the Higgs mass
determination, once realistic event selection criteria for charged leptons and
photons are considered.
\vspace{1pc}
\end{abstract}

\maketitle

\section{Introduction}
Assuming that the Higgs boson will be discovered at the LHC, a measurement of 
its mass, for a large range of values, with a relative experimental precision of 0.1-1\% 
will be possible, by combining ATLAS and CMS measurements and for an integrated luminosity 
of 300~fb$^{-1}$ per experiment~\cite{ATLCMS}. 
To match such an accuracy, the impact of QED radiative 
corrections on the Higgs mass determination through the gluon-fusion process 
$g g \to H \to 4l$ $(4l = 4e, 4\mu, e^+ e^- \mu^+ \mu^-)$ must be evaluated. To this end,
exact ${\cal O}(\alpha)$ and higher-order QED corrections are calculated, 
as described in Sect.~\ref{calc}, and their 
effect on the Higgs mass extraction in the presence of realistic event selection criteria
is evaluated, as discussed in Sect.~\ref{num}. 
Although the electromagnetic corrections affect only the final state, the effects 
of the production process have to be considered, because the typical event selection 
applied on leptons are not Lorentz invariant. The complete process 
$p p \to H \to 4 l(\gamma)$ has been simulated in the narrow width approximation, 
well justified for low Higgs masses, thus allowing to factorize the production 
and decay processes. 
The obtained preliminary results shown in Sect.~\ref{num} indicate that 
QED radiation effects should be carefully considered in view of the expected precision 
at the LHC.

\section{The calculation}
\label{calc}
QED radiative corrections are calculated on top of exact tree-level matrix elements 
for the decay $H \to Z^{(*)} Z^{(*)} \to 4l$, which, at tree level, consists of one diagram
for the final state $e^+ e^- \mu^+ \mu^-$ and two diagrams for equal flavour
leptonic pairs (see Fig.~\ref{diagrammiborn}).

\begin{figure}
\begin{center}
\includegraphics[width=6cm]{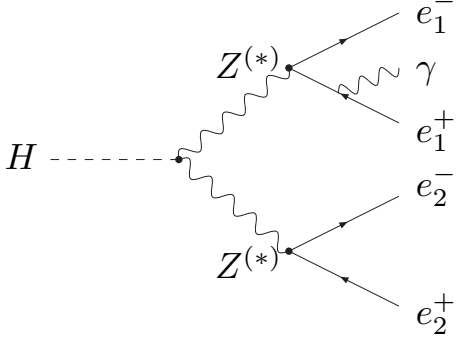}
\caption{The real-radiation diagrams are obtained attaching a real 
photon leg to every final state fermion.}
\label{diagrammireal}
\end{center}
\end{figure}

\begin{figure}[t]
\begin{center}
\includegraphics[width=6cm]{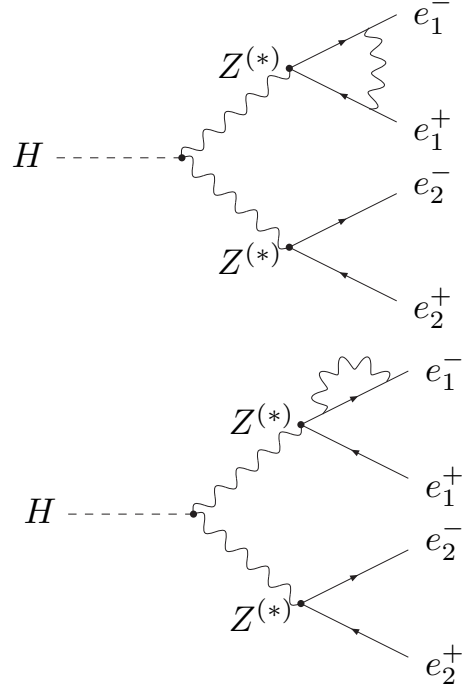}
\caption{Vertex (upper panel) and self-energy (lower panel) diagrams 
contribute to the factorizable corrections.}
\label{diagrammivirtualf}
\end{center}
\end{figure}

\begin{figure}
\begin{center}
\includegraphics[width=6cm]{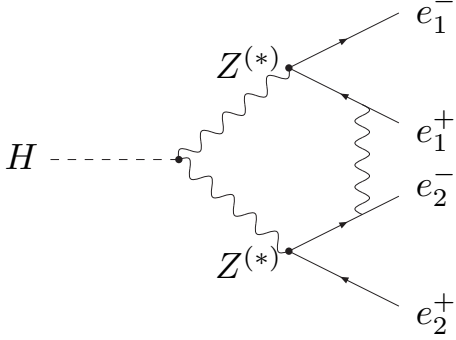}
\caption{Example of pentagon diagram, which contributes to the 
non-factorizable corrections.}
\label{diagrammivirtualnf}
\end{center}
\end{figure}

Since the tree-level is mediated by electrically neutral $Z$-bosons, 
QED corrections can be safely calculated in a gauge invariant way as a subset
of the complete electroweak corrections~\cite{BDDW}, namely neglecting 
$W$-boson exchange contributions. 
Two complementary approaches have been adopted: the former is based on the parton 
shower technique, in the realization of Ref.~\cite{QEDPS}, allowing to calculate 
QED corrections in the leading logarithmic approximation both at ${\cal O}(\alpha)$ and 
to all orders; the latter relies upon a complete ${\cal O}(\alpha)$ perturbative 
calculation. While in the first case the correction is completely factorized over 
the tree-level, in the diagrammatic calculation some care has to be devoted 
to the treatment of the $Z$-boson width in the virtual and real corrections 
in order to safely control the infrared (IR) cancellations. As for any one-loop 
calculation, the expression for the decay width, fully differential over the final 
state, can be written as
\begin{equation}
d\Gamma = (d\Gamma)_B + (d\Gamma)_V + (d\Gamma)_R,
\end{equation}
 where 
the subscripts stand for Born (B), Virtual (V) and Real (R) contributions. 
The IR divergence 
is regularized by means of a small photon mass $\lambda$. The real corrections 
are calculated analytically in soft approximation for $\lambda \leq E_\gamma \leq k_0$
(where $k_0$ is the soft-hard separator), and by means of exact matrix elements 
for $E_\gamma > k_0$, with finite fermion masses and  $\lambda= 0$. The 
real hard photon emission diagrams (one of them is shown in Fig.~\ref{diagrammireal}) 
have been calculated analytically with {\tt FORM}~\cite{FORM} 
and cross-checked with {\tt ALPHA}~\cite{ALPHA}. 
The virtual corrections consist of vertex and self-energy diagrams 
(see Fig.~\ref{diagrammivirtualf}), which 
are symbolically written in terms of Passarino-Veltman form factors~\cite{PV} and 
evaluated numerically with {\tt LoopTools}~\cite{LT}. 
In addition, also box and pentagon diagrams are present. 
One example of five-point graphs is given 
in Fig.~\ref{diagrammivirtualnf}. The pentagon diagrams are reduced, with the 
help of {\tt FORM}~\cite{FORM}, to combinations of four-point form factors 
by means of the techniques introduced in Ref.~\cite{DD}, in order to avoid 
numerical instabilities due to vanishing Gram determinants. The method has 
already been successfully used for the calculation of the ${\cal O}(\alpha)$ 
electroweak corrections to $e^+ e^- \to 4$~fermions, where also six-point 
functions are involved~\cite{DDRW}. Adopting the approximation of vanishing 
fermion masses whenever possible in the virtual corrections and performing 
the calculation in the 't Hooft-Feynman gauge $\xi = 1$, the involved 
five-point functions are at most of rank two and therefore are free 
of ultraviolet divergences. An additional complication is due to the presence of 
the unstable $Z$ bosons. In order to avoid singularities in the phase space, 
the introduction of the $Z$ width is required, which could spoil the 
IR cancellation between virtual and real corrections. In fact the 
IR divergences, contained in the non-factorizable five-point diagrams, are 
cancelled by the interference between real (tree-level) radiation from different 
external legs. A solution is given by the ``complex mass scheme'' 
(introduced in Ref.~\cite{Complex1} for lowest order processes and generalized 
for one-loop calculations in Ref.~\cite{DDRW}), where the $Z$ mass is shifted 
on the complex plane with fixed width $M_Z^2 \to M_Z^2 - i \Gamma_Z M_Z$, both 
in tree-level and in loop diagrams and the couplings become complex quantities, in 
order to respect the Ward identities. 
Considering that self-energies and vertex corrections, neglecting terms of 
${\cal O}(m_f^2/Q^2)$, are already factorized over the tree-level, and that with 
complex $M_Z$ the IR singularity can be factorized over the tree-level also for 
five-point diagrams, the ${\cal O}(\alpha)$ QED corrected Higgs partial width can be 
written as: 

\begin{figure}
\begin{center}
\includegraphics[width=6cm]{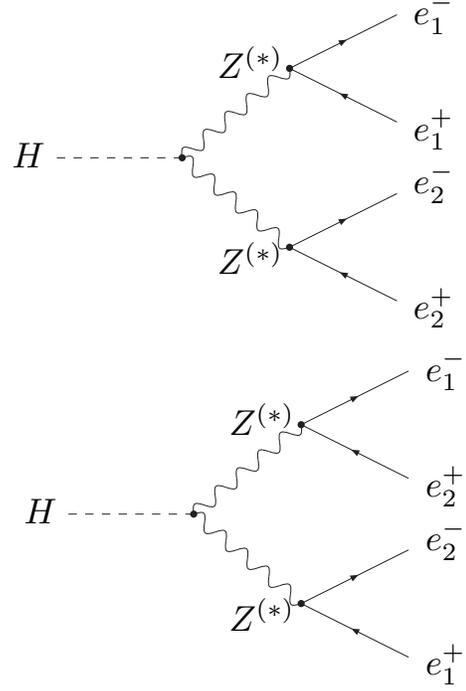}
\caption{Born diagrams for the decay $H\to 4e$.}
\label{diagrammiborn}
\end{center}
\end{figure}

\begin{eqnarray}
d\Gamma_{{\cal O}(\alpha)}&=&(d\Gamma)_B \times (1 + \delta_V^{\rm fact} + \delta_V^{5-IR}) +
(d\Gamma)_R \nonumber \\ 
&&+[(d\Gamma)_V^{\rm nf} - (d\Gamma)_B \times \delta_V^{5-IR}], 
\end{eqnarray}
where $\delta_V^{\rm fact}$ refers to the contribution of self-energies and vertices, 
$\delta_V^{5-IR}$ refers to the IR scalar three-point functions representing the IR part of 
the five-point functions and $(d\Gamma)_V^{\rm nf}$ is the complete contribution of the 
pentagon diagrams. By construction, the IR divergence has been factorized over the 
tree-level, thus allowing for a consistent IR cancellation with the real part $(d\Gamma)_R$, and 
the remainder $[(d\Gamma)_V^{\rm nf} - (d\Gamma)_B \times \delta_V^{5-IR}]$ 
is free of divergence. While the numerical evaluation of two- and three-point scalar 
functions with complex masses is performed with {\tt LoopTools v2.2}, the four-point scalar 
scalar functions is reduced to unidimensional integrals which are evaluated by means 
of the adaptive integration package {\tt CUBA}~\cite{CUBA}. With the same numerical 
algorithm several checks have been performed of the two- and three-point functions 
with complex masses obtained with {\tt LoopTools} (which uses the formulae 
implemented in {\tt FF}~\cite{FF}), finding up-to-digit agreement. 

In order to estimate realistically the impact of QED corrections on the 
Higgs mass determination, the complete 
production and decay process $p p \to H \to 4l+(n \gamma)$ has to be simulated, 
considering typical realistic event selection criteria. Since for a Higgs mass value below 
the real $Z$-pair production threshold the total width is very small compared to 
its mass, the narrow width approximation is adequate. In this approximation 
the calculation can be split in on-shell Higgs 
production $\times$ decay, according to the following formula
\begin{eqnarray}
&&\sigma(pp \to H \to 4l + (n\gamma))=\nonumber\\
&&\int \textrm{d}x_{1}\textrm{d}x_{2}f_{g}(x_{1}, \mu)f_{g}(x_{2},
\mu) \nonumber\\
&&\hat{\sigma}_{gg \to H}\, \delta(x_1x_2s-M_H^2) \nonumber \\
&& {\times \int \frac{\textrm{d} \Gamma(H \to 4l + (n\gamma))}{\Gamma_{tot}}\Theta({\rm cuts})},
\end{eqnarray}
where $f_g(x,\mu)$ are the gluonic parton distribution functions, $\hat{\sigma}_{gg \to H}$ is the parton
level Higgs production cross section and $\Theta({\rm cuts})$ is a step function accounting for the experimental
cuts.

A Monte Carlo 
code has been developed, based on the event generator {\tt ALPGEN}~\cite{ALPGEN} 
for Higgs production and an original library {\tt H24F} for the decay into four leptons with 
QED radiative corrections, taking into account realistic event selections 
for charged leptons and photons. The additional approximation of neglecting 
the Higgs transverse momentum is assumed in the present study. 
A more detailed investigation will be presented elsewhere. 
\begin{figure}[thb]
\includegraphics[width=8.0truecm]{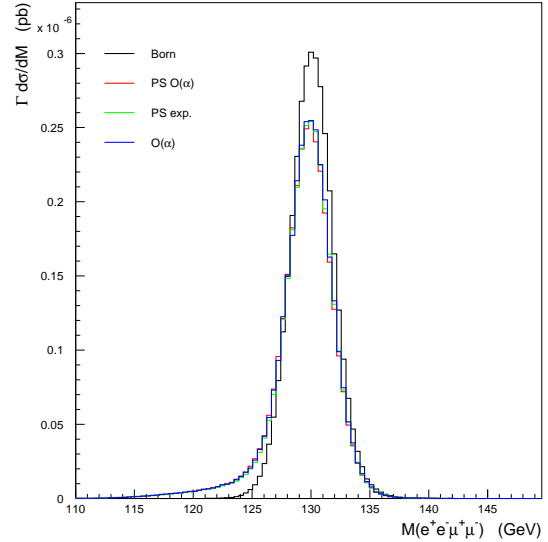}
\caption{Higgs invariant mass in the laboratory frame 
for the decay $H \to e^+ e^- \mu^+ \mu^-$, after applying experimental cuts and momentum
smearing, as obtained at tree level, with Parton Shower and with exact ${\cal O}(\alpha)$ corrections.}
\label{invmasshad}
\end{figure}

\section{Numerical results}
\label{num}
\begin{table*}[thb] 
\caption{The Higgs mass shifts due to 
${\mathcal O}(\alpha)$  and higher-order QED corrections to
$gg \to H \to 4l$ for the three different lepton final states and
for a Higgs mass of 130 GeV.}
\label{hshift}
\begin{center}
\begin{tabular}{|c|c|c|}
\hline
Process & $\vert \Delta (QED)^{(\alpha)}\vert$ 
& $\vert \Delta (QED)^{({\rm exp})} - \Delta (QED)^{(\alpha)} \vert $ \\
\hline
$e^+ e^- e^+ e^-$ & 160 MeV & $\leq$~20 MeV \\
\hline
$e^+ e^- \mu^+ \mu^-$ & 340 MeV & $\leq$~50 MeV \\
\hline
$\mu^+ \mu^- \mu^+ \mu^-$ & 600 MeV & $\sim$~100 MeV  \\
\hline
\end{tabular}
\end{center}
\end{table*}

In order to quantify the shift induced by QED corrections on the Higgs mass
determination, 
binned  $\chi^2$ fits to the (four leptons) invariant mass distribution
(see Fig.~\ref{invmasshad}) have been
performed, following the strategy described in detail in Ref.~\cite{wmass}. A
reference invariant mass distribution is produced at Born level for a given
Higgs mass and $N$ invariant mass distributions are produced including
radiative corrections for different Higgs masses, around the reference value
(with 20 MeV spacing). $N$ $\chi^2$ values are then calculated, according to 
the following equation
\begin{eqnarray}
&&\chi^2(m_H)=\nonumber \\
&&\sum_{i=bins} \frac{\left( \frac{d\sigma_{i,QED}}{\sigma_{QED}}
- \frac{d\sigma_{i,Born}}{\sigma_{Born}} \right)^2}{
\left[ \left( \Delta\frac{d\sigma_{i,QED}}{\sigma_{QED}}\right) ^2
+ \left( \Delta\frac{d\sigma_{i,Born}}{\sigma_{Born}}\right)^2 \right]},
\label{chi2formula}
\end{eqnarray}
where $d\sigma_{i,Born}/\sigma_{Born}$ and 
$d\sigma_{i,QED}/\sigma_{QED}$ are the Born-level and QED corrected cross 
sections relative to the $i$-th bin (normalized to the corresponding integrated
cross sections). The Higgs mass shift is derived from the minimum 
of the $\chi^2$ distribution.

In order to be close to the experimental selection, lepton identification
requirements and kinematical cuts, as well as uncertainties in the energy 
and transverse momentum measurements of the photons/leptons in the detector,
have been implemented according to Ref.~\cite{ATLCMS}. The results of this
preliminary analysis are given in Tab.~\ref{hshift}, showing the Higgs mass
shifts due to  ${\mathcal O}(\alpha)$ (second column) and higher-order 
(third column) corrections, for the three possible four lepton decays and for a
Higgs mass of 130~GeV. It can be 
seen that the mass shift due to multiple photon radiation is, as a rule of
thumb, of the order of 10\% of that caused by one photon emission and,
therefore, non-negligible in view of the
expected precision at the LHC. Furthermore, it has been 
preliminarily observed that, at the present level of accuracy of the
investigation,  exact ${\mathcal O}(\alpha)$ and leading logarithmic 
${\mathcal O}(\alpha)$ QED corrections induce the same mass shifts. A more
detailed analysis, as well as a presentation of further numerical results,
is left to a future work.


\section*{Acknowledgements}
C.M. Carloni Calame is grateful to the Organizers of RADCOR05
for the pleasant and stimulating athmosphere during the conference.

\end{document}